\documentclass{aa}
\usepackage{graphicx,psfig}

\def \fuse{{\it FUSE}}
\def \iue{{\it IUE}}
\def \hst{{\it HST}}
\def \hut{{\it HUT}}

\begin{document}

   \title{Modeling of the Lyman $\gamma$ satellites in \fuse\ spectra
   of DA white dwarfs\thanks{This paper is dedicated in memory of
   J. L. Greenstein, discoverer of the quasi-molecular lines in white
   dwarfs, who passed away on October 21, 2002.} }

   \author{    G. H\'ebrard   \inst{1} 
          \and N. F. Allard   \inst{1,2}
          \and J. F. Kielkopf \inst{3} 
          \and P. Chayer      \inst{4} 
          \and J. Dupuis      \inst{4} 
          \and J. W. Kruk     \inst{4}
          \and I. Hubeny      \inst{5}                 }

   \offprints{G. H\'ebrard (hebrard@iap.fr)}

   \institute {
 Institut d'Astrophysique de Paris, CNRS, 98$^{bis}$ boulevard 
              Arago, F-75014 Paris, France 
  \and  
 Observatoire de Paris-Meudon, LERMA, F-92195 Meudon Principal 
Cedex, France 
  \and 
 Department of Physics, University of Louisville,
         Louisville, KY 40292, USA 
  \and
 Department of Physics and Astronomy, Johns Hopkins University, 
   Baltimore, MD 21218, USA
  \and
 NOAO, 950 North Cherry Avenue, Tucson, AZ 85726,  USA  
  }

   \date{Received 1 April 2003 / Accepted 7 May 2003}

   \abstract{
We present new theoretical calculations of the line profile of
Lyman~$\gamma$ that include transitions in which a photon is absorbed
by a neutral hydrogen atom while it interacts with a proton.  Models
show that two absorption features located near 992~\AA\ and 996~\AA\ 
are due to H-H$^+$ collisions.  These quasi-molecular satellites are
similar to those that were identified in the wings of Lyman~$\alpha$
and Lyman~$\beta$ lines of hydrogen-rich white dwarfs.  We compute
synthetic spectra that take account of these new theoretical profiles
and compare them to the spectra of four DA white dwarfs that were
observed with \fuse.  The models predict the absorption features that
are observed in the wing of Lyman~$\gamma$ near 995~\AA, and confirm
that these features are quasi-molecular satellites.

   }

\authorrunning{G. H\'ebrard et al.}
\titlerunning{Modeling of the Lyman $\gamma$ satellites in 
               \fuse\ spectra of DA white dwarfs}

   \maketitle
%

\section{Introduction}

Quasi-molecular satellites are absorption features due to transitions
that take place during close collisions of a radiating hydrogen atom
with a perturbing atom or proton.  These spectral features, which are
present in the red wings of the Lyman series lines, provide an
important source of opacity in the atmospheres of hydrogen-rich white
dwarfs.

The {\it IUE} spectra of the DA white dwarf 40~Eri~B were the first
ones which revealed a strong, unexpected absorption feature near
1400~\AA\ (Greenstein 1980). Absorption features near 1600~\AA\ were
thereafter detected in spectra of cooler DA white dwarfs (Holm et
al.~1985).  These features remained unexplained until Koester et
al.~(1985) and Nelan \& Wegner (1985) simultaneously identified them
as quasi-molecular satellites of the Lyman~$\alpha$ line. The H$_2$ and
H$_2^+$ satellites at 1600~\AA\ and 1400~\AA\ were thereafter observed
in other DA white dwarfs, as well as in laboratory plasmas (Kielkopf
\&\ Allard 1995).  The H$_2^+$ satellite absorption features at
1058~\AA\ and 1076~\AA\ were first identified in the spectrum of the
DA white dwarf Wolf$\,$1346, as observed with {\it HUT} (Koester~et
al.~1996).  {\it ORFEUS} (Koester et al.~1998) and \fuse\ (Wolff et
al.~2001; H\'ebrard et al.~2002b) observations allow these
Lyman~$\beta$ satellites to be observed in other targets.

New calculations for the absorption profiles of Lyman~$\gamma$ line of
atomic hydrogen perturbed by protons have been used to predict
synthetic spectra of hot hydrogen-rich white dwarfs and allow the
identification of a strong feature near 995~\AA.  This feature was
present in the {\it HUT} spectrum of \object{Wolf$\,1346$} (Koester et
al.~1996) but it is only in Koester et al.~(1998) that a similar
feature present in {\it ORFEUS} spectra of WD$\,$1031$-$114 and
WD$\,$0644$+$375 was suspected to be a new satellite in the wing of
Lyman~$\gamma$.  It was also distinctly detected in \fuse\ spectra of
\object{CD$\,-38^{\circ}10980$} (Wolff~et al.~2001) and of 
Sirius~B (Holberg~2002).

These last two recent observations provided motivation to investigate
the theory of the Lyman~$\gamma$ line profile using accurate
theoretical molecular potentials, instead of using the simple Stark
broadening approximation. Our new calculations of Lyman~$\gamma$
satellites are summarized in Sect.~2.  We present in Sect.~3 a
comparison between synthetic spectra including Lyman~$\gamma$
satellites and the \fuse\ spectra of four DA white dwarfs.  The
results are discussed in Sect.~4.

\section{Theoretical line profiles} 

The theoretical approach is based on the theory of pressure broadening
due to Baranger~(1958), developed in an ``adiabatic representation''
that does not exclude degeneracy of atomic levels.  A detailed
description of our semi-classical approach as applied to the shape of
the Lyman lines has been given by Allard~et al.~(1994) and Allard et
al.~(1999).
 
The Lyman~$\gamma$ profile and its satellites are calculated for
physical conditions encountered in the atmospheres of white dwarfs.
Because the density of the hydrogen atoms is low (10$^{15}$ to
10$^{17}$~cm$^{-3}$), we computed the line profile by using the low
density approximation as described by Allard et al. (1994).  This
approximation uses the expansion of the autocorrelation function in
powers of density.  We also used accurate theoretical potentials
describing the binary interactions of one hydrogen atom with a ionized
hydrogen atom.  The H$_2^+$ molecular potentials are available for the
H~($n \le 3$) states (Madsen \& Peek 1971).  We computed all those
related to $n=4$ using a code kindly provided by J.~Peek.

Our theoretical approach allows the variation of the radiative dipole
moment during a collision.  But through lack of appropriate data for
Lyman~$\gamma$, we assumed that the dipole moment for the transitions
is constant.  Such approximation has been applied to the calculations
of the quasi-molecular satellites observed in the \iue\ and \hst\
spectra of white dwarfs (see, {\it e.g.}, Koester \& Allard 1993;
Koester et al. 1994; Bergeron et al. 1995).  In the last section we
discuss the possible effects of a variable dipole moment on the line
profile calculations.

\begin{figure}
\centering
{\resizebox{0.48\textwidth}{!}{ \includegraphics*{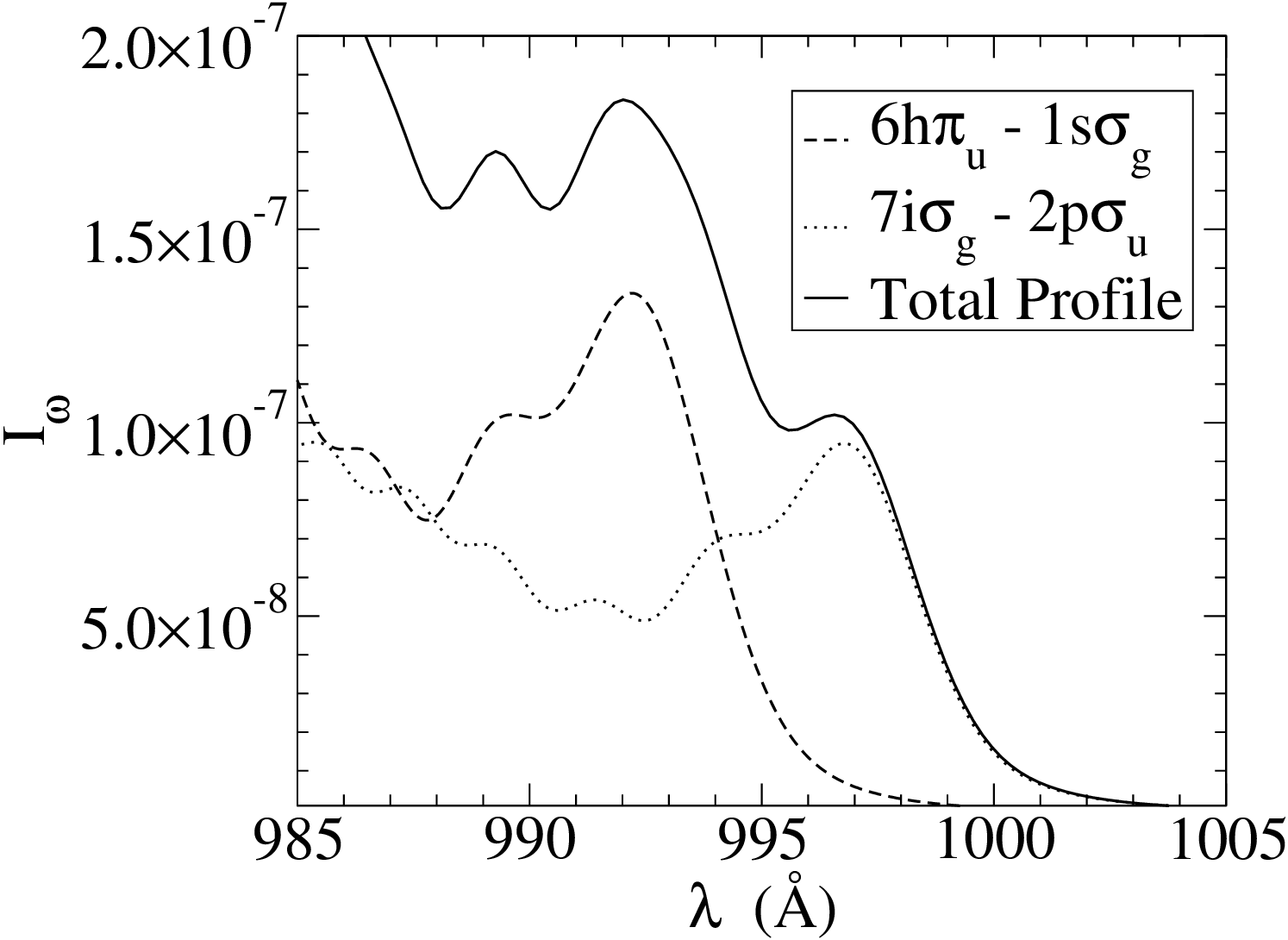}}}
\caption {The 995~\AA-region Lyman~$\gamma$ wing.
The individual contribution of the two transitions that 
give rise to the two main satellites are over-plotted. 
$I_\omega$ is the normalized line profile proportional to the
absorption coefficient as described by Allard et al.~(1999). 
}
\label{satgamma}
\end{figure}

An H$_2^+$ correlation diagram was constructed for Lyman~$\gamma$
(Allard~et al. 2003), to which 14 transitions contribute.  Two of
these transitions, $ 1s \, \sigma_{g} \rightarrow 6h \,\pi_{u}$ and $
2p \, \sigma_{u} \rightarrow 7i \, \sigma_{g}$, yield satellites in
the red portion of Lyman~$\gamma$ at 992~\AA\ and 996~\AA.
Fig.~\ref{satgamma} shows the profiles of both individual satellites
and their effect on the total profile of Lyman~$\gamma$.  We obtain a
blend of the line satellites with a maximum near 992~\AA. The profile
gives a shape similar to the 995~\AA\ feature observed in white dwarf
spectra.

\section {Astronomical applications to synthetic spectra for 
DA white dwarfs}

Atmosphere models have been calculated using the programs TLUSTY
(Hubeny 1988; Hubeny \& Lanz 1992, 1995).  We considered LTE model
atmospheres with pure hydrogen composition that explicitly include the
Lyman~$\alpha$ and Lyman~$\beta$ quasi-molecular opacities.  The
resulting spectra were computed by using the spectral synthesis code
SYNSPEC that incorporates the quasi-molecular satellites of
Lyman~$\alpha$, Lyman~$\beta$, and Lyman~$\gamma$. Models predict that
the Lyman~$\gamma$ satellites are visible roughly for effective
temperatures between $15\,000$~K and $30\,000$~K.  We verified that
non-LTE effects are unimportant in the range of effective temperatures
considered in this study.

\begin{small}
\begin{table*}
\caption{Targets.}
\vspace{0.2cm}
\begin{tabular}{lcccccccc}
\hline
\hline  
Target    & Observation & Obs. Date & 
Mode & Slit & Exp. (ks) & CalFUSE$^{\rm a}$ & $T_{\rm eff}$ &$\log g$\\
\hline
\hline  
Wolf$\,1346$        & B1190501    & Jul 2001 & TTAG & LWRS &
5.8  &  1.8.7 & $20\,000$ & 8.0 \\
\hline
EG$\,102$          & B1190301    & May 2001 & TTAG & LWRS &
10.7 &  1.8.7 & $20\,435$      &  7.9    \\
                  & S6010401    & Jan 2002 & TTAG & LWRS & 17.4 & 
1.8.7   &      \\
\hline
BPM$\,6502$        & Z9104501    & Jun 2002 & TTAG & LWRS &
32.0 & 1.8.7 & $21\,380$       &  7.9    \\
\hline
CD$\,-38^{\circ}10980$        & Q1100101    & Jul 2000 &
 HIST & MDRS & 4.8  &  2.2.1 & $23\,000$     &   7.9   \\
\hline
\end{tabular}
\\
$^{\rm a}$ Version of the pipeline used for spectral extraction. 
\end{table*} 
\end{small}

We used observations of four DA white dwarfs performed with \fuse\ 
(Moos et al.~2000).  The spectra were obtained between 2000 and 2002
and were retrieved from the \fuse\ public database (see Table~1).  The
one-dimensional spectra were extracted from the two-dimensional
detector images and calibrated using the CalFUSE pipeline.  When
possible, we did not use exposures with strong airglow emission.  We
also excluded regions of the spectra near the edges of the detector
segments, which provide poor quality data.  The eight \fuse\ detector
segments of the different exposures were co-added together and
projected on a $0.16$~\AA-pixel base, so pixels about 25 times larger
than the original \fuse\ detector pixels.  This degradation of the
\fuse\ spectral resolution (typically
$\lambda/\Delta\lambda\simeq15000$ for this kind of target with the
slit LWRS or MDRS; see H\'ebrard et al. 2002a or Wood et al. 2002) has
no effect on the shapes of the large stellar features which we study
and allows the signal-to-noise ratio to be increased.

In Fig.~2 we compare the \fuse\ spectra with the models that we
obtained for these targets.  The effective temperatures are within the
intermediate temperature range for the visibility of both the
Lyman~$\beta$ and Lyman~$\gamma$ satellites.  The agreement indicates,
by the coincidence of position and approximate shape, that the large
feature near 995~\AA\ is indeed a satellite of Lyman~$\gamma$ due to
the blend of two close features at 992~\AA\ and 996~\AA.

\begin{figure*}
\centering
{\resizebox{\textwidth}{!}{ \includegraphics*{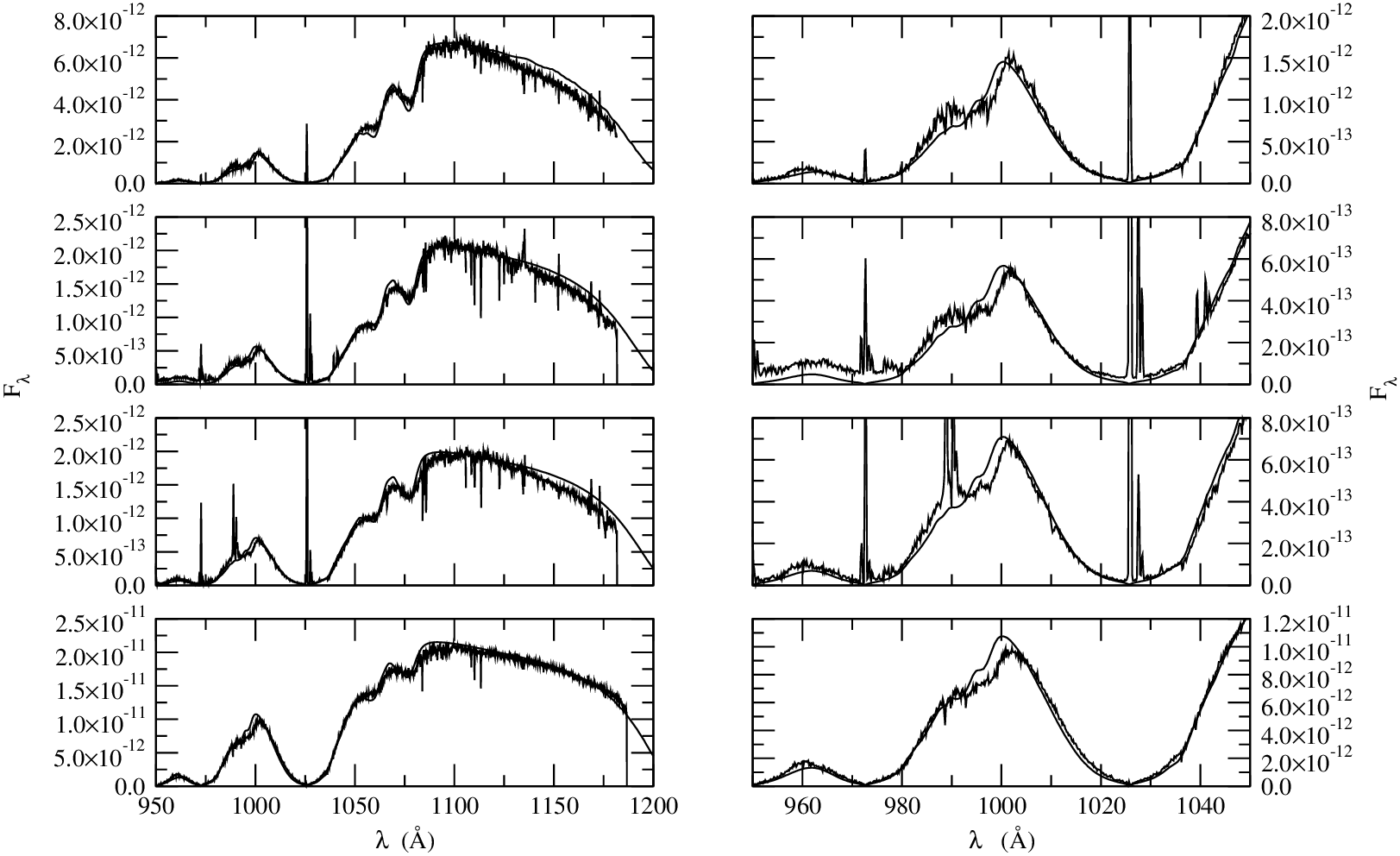}}}
\caption {A comparison of the \fuse\ spectra of the  white dwarfs 
and the theoretical models.
Top to bottom: Wolf$\,1346$, EG$\,102$, BPM$\,6502$, and 
CD$\,-38^{\circ}10980$. The flux is in erg/cm$^2$/s/\AA. 
The Lyman~$\beta$ satellites are at 1058~\AA\ and 1076~\AA; 
the Lyman~$\gamma$ satellites (enlarged in right panel) 
are at 992~\AA\ and 996~\AA. The narrow emission features
are due to terrestrial airglow.}
\label{wdprofs}
\end{figure*}

The first target plotted in Fig.~2 is Wolf$\,1346$.  The comparison
for the \fuse\ spectrum shows that the far UV is well fitted with a
synthetic spectrum computed with our new profile calculations for
$\log g=8.0$ and $T_{\rm eff}=20\,000$~K. These atmospheric parameters
are close to those determined from optical spectra (7.9; 20\,000) and
adopted by Koester~et al.~(1996) to fit the {\it HUT} spectrum of that
white dwarf.  The Lyman~$\gamma$ satellite, already observed in the
\hut\ spectrum, is well-reproduced by our synthetic spectrum.

The next two objects, \object{EG$\,102$} and \object{BPM$\,6502$}, are
slightly hotter white dwarfs, with smaller gravities.  We used the
effective temperatures and gravities determined from optical spectra
(Finley~et al.~1997; Bragaglia et al.~1995).  The predicted spectra
for EG$\,102$, with $T_{\rm eff}=20\,435$~K and $\log g=7.87$, and for
BPM$\,6502$, with $T_{\rm eff}=21\,380$~K and $\log g=7.86$, fit quite
well the \fuse\ spectra.  In the case of BPM$\,6502$, all the
\fuse\ exposures present strong airglow emissions, so it was not
possible to remove them. These two stars are not normal DA white 
dwarfs: EG$\,102$ presents metallic lines (Holberg et al.~1997) 
and BPM$\,6502$ is in a post common-envelope binary system 
(Kawka et al.~2000, 2002). These particularities seem however to have 
no effects on the quality of the fits. 

The hottest object we present is CD$\,-38^{\circ}10980$. The \fuse\ 
observation that we used is the same as the one presented by Wolff~et
al.~(2001), but our data reduction was performed with a more recent
version of CalFUSE. Our best fit was obtained using $T_{\rm
eff}=23\,000$~K and $\log g=7.9$. The values of the atmospheric
parameters are slightly smaller than the ones determined from optical
spectra (Vauclair et al.~1997) to get a better agreement with the
Lyman~$\gamma$ satellite.

CD$\,-38^{\circ}10980$ is near the upper limit of the expected
visibility range of the Lyman~$\beta$ satellites.  Nevertheless,
Lyman~$\beta$ and Lyman~$\gamma$ quasi-molecular satellites have been
observed in the \fuse\ spectrum of the ultra-massive white dwarf
PG$\,1658+441$ ($T_{\rm eff} = 30\,500$K, $\log g = 9.36$) as reported
by Dupuis et al.~(2001). We speculate that the presence of
quasi-molecular satellites in this star is related to its high surface
gravity and we will investigate the question in more details in Dupuis
et al.~(2003). It is interesting to note that Sirius~B ($T_{\rm eff} =
24\,790$K, $\log g = 8.57$; Holberg et al.~1998), which also has a
relatively high surface gravity, shows the quasi-molecular satellites
(Holberg~2002).

Finally, the four spectra show a kink in the profile around 1037~\AA\ 
that is well reproduced by our models (Fig.~2). This feature is
probably the first identification of the H$_2^+$ satellite of
Lyman~$\beta$, as predicted in Fig.~7 of Allard et al. (1998a.)

\section{Conclusion}

The good agreement between the \fuse\ spectra and our calculations
allowed the line satellites near 995~\AA\ to be identified, although
there are still some systematic discrepancies between the predicted
and observed profiles. The shape of the profile in the region of the
satellites is sensitive to the relative strength of these two main
features.  It is therefore important to get an accurate quantitative
determination of the satellite amplitudes.  Accurate theoretical
molecular potentials have to be used to describe the interaction
between the radiator and the perturber as we did in the work described
here.  Another important factor, not yet included, is the variation of
the dipole moment during the collision.  Allard~et al.~(1998a, 1998b,
1999) have shown that the strengths of line satellites are dependent
on values of the electric-dipole moments at the internuclear
separation responsible for the satellites. Large enhancements in the
amplitudes of the satellites may occur whenever the dipole moment
increases through the region of internuclear distance where the
satellites are formed. The theoretical shape in this region of the
profile is then dependent on the dipole moment of the
transition. Some of the differences between the observed and the 
synthetic spectra may be due to our constant dipole moment
approximation. That will be improved in future calculations.

\begin{figure}  
\centering    
\resizebox{0.48\textwidth}{!}{\includegraphics*{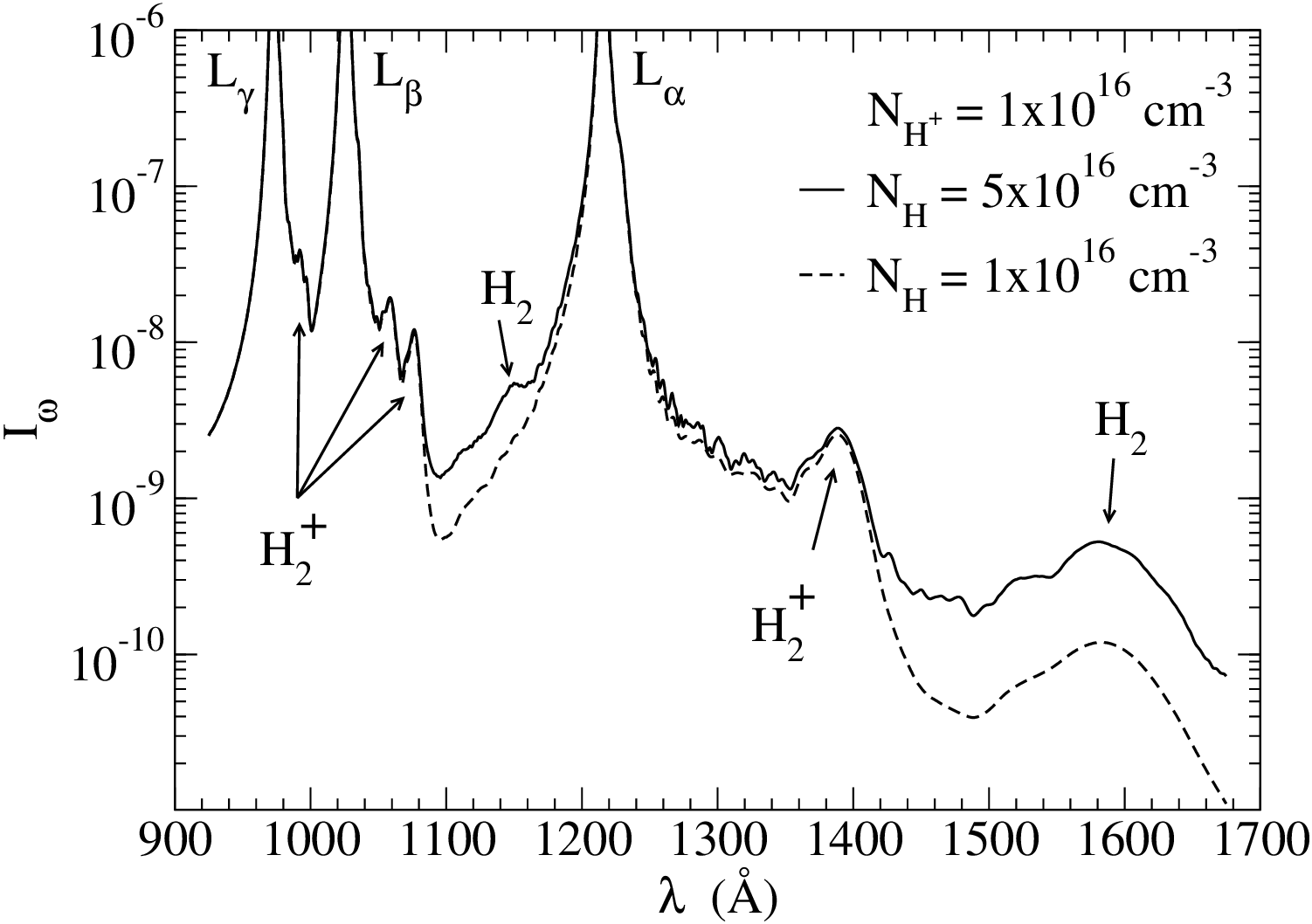}}
\caption{Total profile of Lyman~$\alpha$, Lyman~$\beta$, and 
Lyman~$\gamma$ perturbed by neutral hydrogen
and protons. Two different neutral densities ($1\times10^{16}$~cm$^{-3}$ 
and $5\times10^{16}$~cm$^{-3}$) are compared for a fixed
ion density ($1\times10^{16}$ cm$^{-3}$), and $T=25\,000$~K. 
Lyman~$\gamma$ profile is only perturbed by protons.}
\label{lytotal}
\end{figure}

As a summary we present in Fig.~\ref{lytotal} the sum of the profiles
of Lyman~$\alpha$, Lyman~$\beta$, and Lyman~$\gamma$ perturbed by
collisions with neutral hydrogen and protons for two different
densities of neutral hydrogen.  We can note that beside the
quasi-molecular lines already detected, there is another predicted
H$_2$ satellite at 1150~\AA\ (Allard~et al.~2000), in the red wing of
Lyman~$\beta$, which should be observed in cool DA white dwarfs in the
range of effective temperatures (11000 to 13000~K), where the variable
ZZ~Ceti objects are found (H\'ebrard~et al.~2002b). This line
satellite has never been observed in stars or laboratory plasmas.
Future \fuse\ observations and ongoing study of the FUV spectrum of
dense hydrogen plasmas may allow this feature to be detected.

\begin{acknowledgements}
This work is based on data obtained by the NASA-CNES-CSA \fuse\ 
mission operated by the Johns Hopkins University. Financial support to
U. S.  participants has been provided by NASA contract NAS5-32985.
French participants are supported by CNES.  The molecular potential
computation code was kindly provided by J.~Peek; we would like to
thank him.

\end{acknowledgements}

\end{document}